# Semiconductor-Dielectric-Metal Solar Absorbers with High Spectral Selectivity

Eric J. Tervo[1,*] and Myles A. Steiner[1]

[1]National Renewable Energy Laboratory, Golden, CO 80401
*eric.tervo@nrel.gov

An ideal solar thermal absorber has a sharp transition between high and low absorptance at the wavelength where the blackbody emissive power begins to exceed the solar irradiance. However, most real selective absorbers have a fairly broad transition, leading to both solar absorption and thermal emission losses. Here, we model, fabricate, and characterize a highly selective semiconductor-dielectric-metal ($Ga_{0.46}In_{0.54}As$ - $MgF_2$ - Ag) solar absorber with an extremely sharp transition from high to low absorptance. The thin semiconductor serves as a selective filter, absorbing photons with wavelengths shorter than the bandgap and transmitting those with longer wavelengths. The highly reflective dielectric-metal rear mirror allows the structure to have very low emittance for longer wavelengths. These characteristics provide the absorber with a measured solar absorptance >91% below the bandgap wavelength and infrared emittance <5% at 100 °C above the bandgap wavelength. This transition wavelength can be tuned by modifying the semiconductor composition, and modeling indicates that the absorber's optical properties should be stable at high temperatures, making the structure a good candidate for unconcentrated to highly concentrated solar thermal energy conversion.

## Introduction

Solar thermal energy can be used to drive a variety of industrial processes, reducing their demand for electricity and fossil fuels [1, 2]. It also can provide heat for concentrating solar power plants, in which concentrated sunlight is absorbed to heat a fluid or thermal energy storage medium. This solar heat is then used to drive a turbine and produce electricity. When a concentrating solar power plant has integrated thermal energy storage, it can be both an efficient and dispatchable source of renewable energy [3, 4]. Regardless of the application, solar thermal energy converters must absorb as much solar radiation as possible while emitting as little thermal radiation as possible, the latter representing a direct efficiency loss. This motivates the use of a selective solar absorber [3, 5-12], as illustrated for a parabolic trough collector in Fig. 1a. Selective absorbers maximize spectral absorptance at short wavelengths where the solar radiation flux exceeds the absorber's blackbody emissive power, and they minimize absorptance for long wavelengths where the emissive power is higher, as shown in Fig. 1b for an ideal absorber under a solar concentration of 100 and operating at 500 °C. The wavelength at which the solar spectrum and the emission spectrum cross each other can vary depending on the temperature of the absorber and the intensity of the sunlight, and so the ideal absorber also depends on its particular operating conditions.

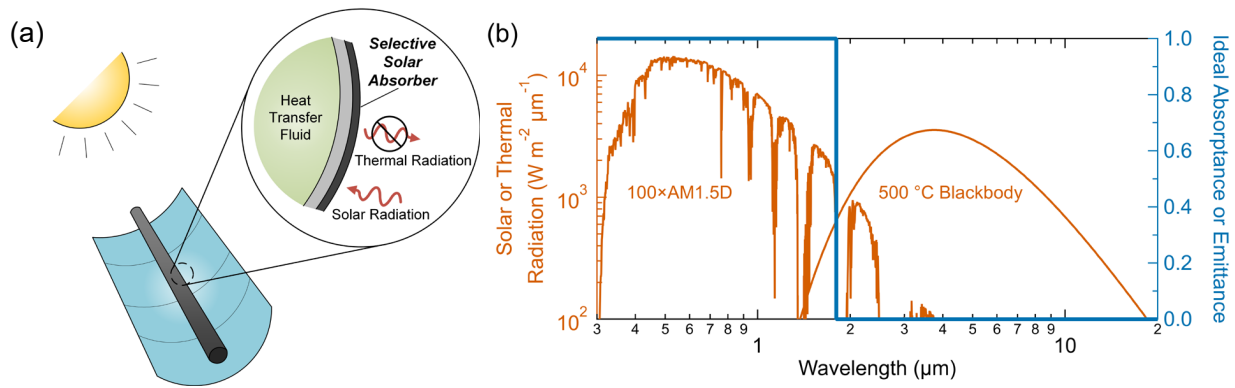

**Figure 1.** (a) Simplified schematic of a parabolic trough collector for concentrating solar power showing the role of a selective solar absorber: to absorb as much sunlight as possible while emitting as little thermal radiation as possible. (b) For the AM1.5D solar spectrum under 100x concentration and an operating temperature of 500 °C (which is in the expected range for a parabolic trough collector [3]), the ideal spectral absorptance/emittance has an abrupt transition from high to low where the blackbody spectrum exceeds the solar spectrum.

Selective solar absorbers generally fall into six categories [3, 6]: (1) intrinsic, which are composed of a homogeneous material that has inherent spectral selectivity; (2) semiconductor-metal tandems, which rely on the semiconductor to absorb solar photons with energy above the bandgap and the metal to provide low infrared emittance; (3) multilayer thin-films, which use constructive and destructive interference effects to tailor the absorptance characteristics; (4) cermets, which are metal-ceramic composites that act as a graded index layer on a reflective substrate; (5) structured surfaces, which use microstructured geometry to trap short wavelength light while reflecting longer wavelengths; and (6) photonic crystals, which utilize periodic nano- to micro-scale features for photonic band engineering. All of these have their own advantages and disadvantages. For example, multilayer thin-films and photonic crystals allow for very fine control over the absorption characteristics, but their small features may lead to challenges with stability at high temperatures [7]. Cermet absorbers have seen the most commercial success, in large part due to their low cost and good high temperature stability in vacuum, but they do not typically provide high spectral selectivity [9]. The different types of selective absorbers, their history, and their advantages and disadvantages are reviewed in detail elsewhere [3, 5-12].

Semiconductor-metal tandem absorbers are particularly interesting because they can exhibit many traits of a high-performance absorber, including high spectral selectivity, high temperature stability, and a relatively simple structure that is amenable to various manufacturing methods. Since they were proposed in the early 1960s [13], many different semiconductor materials for solar absorbers have been investigated, including PbS [14-16], Si [17-21], Ge [18, 21, 22], and $Si_{0.8}Ge_{0.2}$ [23]. Early work by McMahon and Stierwalt with PbS on Al [15] was promising, reaching a solar absorptance of 0.90 and infrared emittance of just 0.038 at 258 °C. However, the relatively long transition wavelength of PbS near 3 μm is only well-suited for low temperature absorbers approximately less than 100 °C, is nonideal for higher temperature applications (see Fig. 1), and other research groups working with PbS have not achieved the same level of performance [14, 16]. More recent work with Ge by Thomas *et al.* [22] and with $Si_{0.8}Ge_{0.2}$ by Moon *et al.* [23] show favorable transition wavelengths in the 1.2 – 1.8 μm range, but they are not able to achieve simultaneously high solar absorptance and low infrared emittance (solar absorptances of 0.76 and 0.90-0.95 and infrared emittances of 0.05 and ~0.30 for Ge and $Si_{0.8}Ge_{0.2}$, respectively). Modeling indicates that semiconductor-metal absorbers could perform much better, with solar absorptances >0.90 and infrared emittances <0.10 for temperatures exceeding 700 °C [21], which motivates further efforts to design and fabricate these types of absorbers.

In this work, we design, fabricate, and characterize a semiconductor-based selective absorber that incorporates two novel features to improve its performance. First, we use the ternary III-V semiconductor $Ga_{0.46}In_{0.54}As$, whose composition may be altered to tune the transition wavelength between 0.87 μm (GaAs) and 3.50 μm (InAs) at room temperature. This allows our general structure to be refined for a particular application's combination of solar concentration and absorber temperature in order to optimize its performance. Second, we incorporate a thin $MgF_2$ dielectric layer in between the semiconductor and the rear metal reflector to boost the infrared reflectance and thus lower the infrared emittance. This is a concept borrowed from recent work on thermophotovoltaics [24-27], which has shown that the dielectric layer reduces infrared losses caused by waveguide modes that can exist at a semiconductor-metal interface. Taken together, these features give our semiconductor-dielectric-metal absorber a very sharp transition from high to low absorptance at a tailored wavelength and simultaneously maintain high solar absorptance and low thermal emittance.

**Methods**

<u>Absorber Fabrication</u>

A schematic of our semiconductor-dielectric-metal selective solar absorber is shown in Fig. 2a, a cross-sectional scanning electron microscope image of the structure is shown in Fig. 2b, and a photograph of the absorber is shown in Fig. 2c. The absorbing semiconductor layer is a thin film of $Ga_{0.46}In_{0.54}As$ grown by atmospheric pressure metalorganic vapor phase epitaxy on an InP substrate. The material sources are trimethylgallum and trimethylindium for the group-III elements, and arsine and phosphine for the group Vs. The substrate is an (001)-oriented, sulfur-doped InP wafer, miscut 2° toward the <110> direction, with a 2" diameter. The wafer is loaded into the reactor and heated to 620°C under a mixture of phosphine and purified hydrogen, flowing at ~6 L/min. The surface is deoxidized for one minute at 620°C under a phosphine overpressure, followed by growth of a 0.1 μm InP seed layer, a pair of etch-stop processing layers including 1 μm of GaInAs and 0.5 μm of InP, and then the thick GaInAs absorber layer. The sample is cooled back to room temperature under an arsine overpressure. The absorber layer is grown at a nominal rate of 6.8 μm/hr and a V/III ratio of 30.

After growth, the rear mirror consisting of $MgF_2$, 1 nm of Ti for adhesion (not shown in Fig. 2), and Ag is deposited in vacuum by thermal evaporation ($MgF_2$) and electron beam evaporation (Ti and Ag) on the epitaxial surface. The sample is then bonded to a piece of an intrinsic Si wafer, which serves as a mechanical handle, by thermally conductive epoxy between the Ag and the Si wafer. This step leaves the absorber with the substrate-side up, and the substrate is etched away in HCl : $H_3PO_4$ (4:1). The two etch-stop layers are removed with $H_3PO_4$ : $H_2O_2$ : $H_2O$ (3:4:1) and HCl. Finally, an antireflective coating consisting of four alternating thin layers of ZnS and $MgF_2$ is deposited in vacuum by thermal evaporation. Some of these layer choices were made based on available facilities, but they could be replaced with more durable options, such as a wafer bond instead of epoxy and a three-layer oxide antireflective coating instead of the four-layer $MgF_2$/ZnS coating.

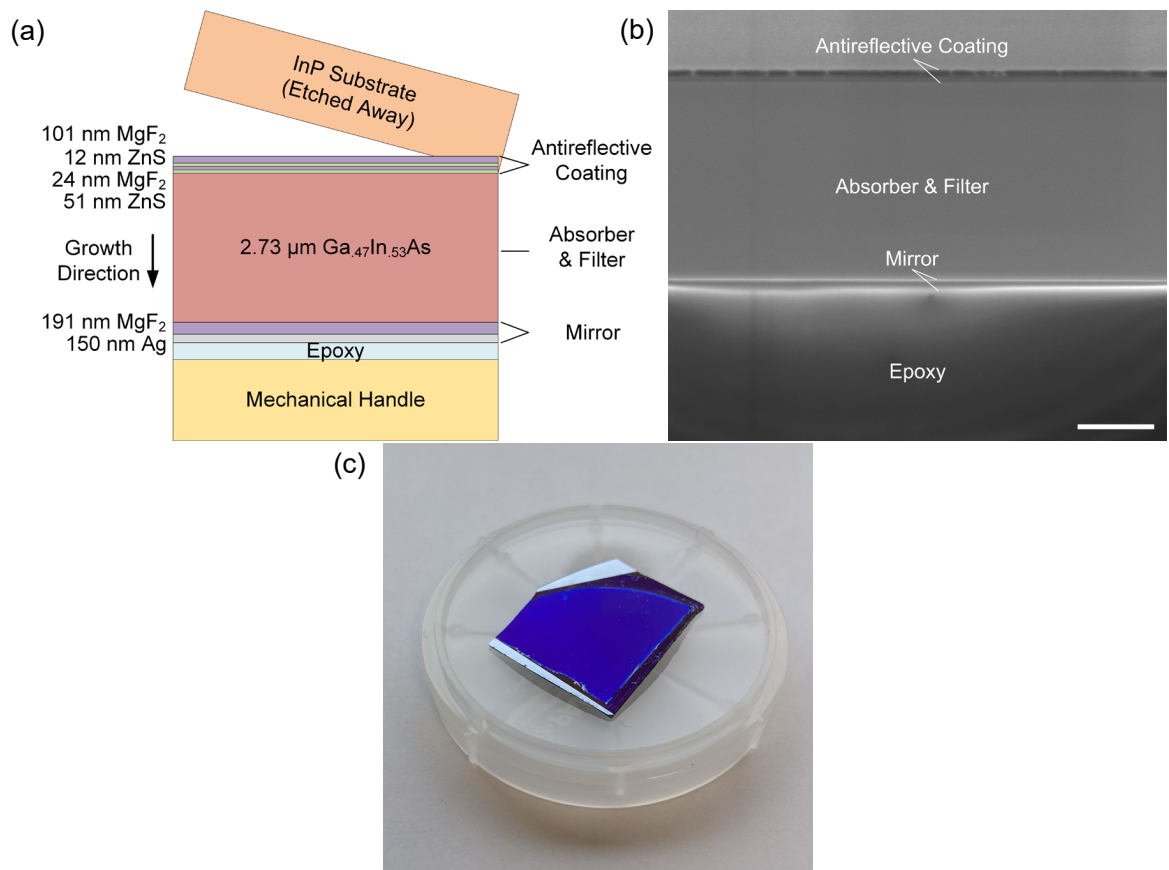

**Figure 2.** (a) Schematic with nominal thicknesses, (b) scanning electron microscope (scale bar is 1 µm), and (c) photograph images of the selective absorber. The sample is grown and fabricated via inverted semiconductor processing, where the GaInAs layer is first grown upside down, the rear mirror layers are then deposited and the device is glued to a silicon substrate, the growth substrate is removed by chemical etching, and finally the antireflective coating layers are deposited.

Absorber Characterization

The ultraviolet, visible, and near-infrared (wavelengths of 300 nm – 2.5 µm) reflectance was measured with an Agilent Cary 7000 spectrophotometer with an Agilent external diffuse reflectance accessory, which collects all angles of reflection *via* an integrating sphere from an 8° angle of incidence. The absorber or reflectance standard was mounted at the reflection port of the diffuse reflectance accessory on a Linkam HFS600E-PB4 stage to vary the sample temperature between 30 °C and 100 °C. This temperature range is used because the epoxy will melt at temperatures higher than 100 °C, but a different bonding mechanism would allow for much higher temperatures to be tested. All measurements were referenced to a NIST-traceable Al reflectance standard.

The infrared (wavelengths of 2.5 µm – 20 µm) reflectance was measured with a SOC-100 Hemispherical Directional Reflectometer coupled to a Nicolet 6700 FTIR spectrometer. The 10° angle of reflectance was measured from hemispherical illumination, which by reciprocity is equivalent to the hemispherical reflectance from 10° angle of incidence. A temperature-controlled sample mounting stage within the SOC-100 was used to vary the sample temperature, and all measurements were referenced to a specular Au standard provided by the instrument manufacturer.

Absorber Model

The spectral absorptance and emittance of our selective absorber structure was calculated using the standard transfer matrix method [28], which requires the optical properties of each layer as an input. Tabulated optical constants were used for Ag [29], Ti [30], MgF$_2$ [31], and ZnS [32]. The optical properties of the semiconductor, however, may exhibit significant variation with temperature and composition in the form of a shift in the bandgap wavelength and/or free carrier emission. To account for these effects, the real part of the relative permittivity $\varepsilon = \varepsilon' + i\varepsilon''$ of Ga$_x$In$_{1-x}$As was calculated according to the critical-point parabolic-band model by Adachi [33]:

$$\varepsilon'(E) = \varepsilon'_{Adachi}(E) \quad (1)$$

where $E$ is the photon energy. The imaginary part was calculated with the same model for energies above the bandgap (wavelengths shorter than the bandgap):

$$\varepsilon''(E > E_g + 5 \text{ meV}) = \varepsilon''_{Adachi}(E) \quad (2)$$

where $E_g$ is the bandgap energy. All input parameters to Adachi's model are the same as those specified in reference [33] except for the strength of the band edge transition $A$, which is fit to experimental data as $A = 2.0$ eV$^{3/2}$, and the bandgap energy. The temperature-dependent bandgap energy is calculated according to Zielinski *et al.* as [34]

$$E_g(T) = E_g(0) - \frac{B(\hbar\Omega)^{1/2}}{e^{\hbar\Omega/(kT)} - 1} \quad (3)$$

where $B = 17.6$ meV$^{1/2}$ and $\hbar\Omega = 18.9$ meV are fitting parameters from reference [34] and $E_g(0) = 800$ meV is the bandgap at 0 K fit to our data. The 300 K bandgap from Eqn. (3) is used with the bowing parameters from reference [35] to extract the composition of the Ga$_x$In$_{1-x}$As for a given value of $E_g(0)$. Note that the composition of Ga$_{0.46}$In$_{0.54}$As as determined here from the bandgap is with 1% of the composition of Ga$_{0.47}$In$_{0.53}$As for the lattice-matched alloy as determined from Vegard's law [36] of linear variation of composition with lattice constant; this accuracy is sufficient for our purposes. Near the bandgap, we use the approach of Guerra *et al.* [37] for the imaginary part to more accurately model the absorption edge and the Urbach tail. For lower energies (longer wavelengths), we include free-carrier effects by also incorporating a Drude model [38]:

$$\varepsilon''(E \leq E_g + 5 \text{ meV}) = \varepsilon''_{Guerra}(E) + \varepsilon''_{Drude}(E) \quad (4)$$

We define the first term on the right side of Eqn. (4) as [37]

$$\varepsilon''_{Guerra}(E) = -\frac{\varepsilon''_c}{2}\sqrt{\pi E_U}\text{Li}_{1/2}\left(-e^{\frac{E-E_g}{E_U}}\right) \quad (5)$$

where $\varepsilon''_c = 4.4$ and $E_U = 5$ meV are fitting parameters representing an imaginary permittivity and the Urbach energy and $\text{Li}_{1/2}$ is the Polylogarithm of order 1/2. The second term on the right side of Eqn. (4) is [38]

$$\varepsilon''_{Drude}(E) = \text{Imag}\left(\varepsilon_\infty - \frac{E_p^2}{E^2 + iE\hbar\gamma}\right) \quad (6)$$

where $\hbar$ is the reduced Planck constant; $\varepsilon_\infty$ is the high-frequency relative permittivity; $E_p = \hbar\sqrt{Ne^2/m_n^*\varepsilon_0}$ is the plasma energy with the carrier concentration $N$, the electron charge $e$, the carrier effective mass $m_n^*$, and the permittivity of free space $\varepsilon_0$; and $\gamma = e/m_n^*\mu$ is the scattering rate with the mobility $\mu$. Because our $Ga_{0.46}In_{0.54}As$ is grown without dopants, we take $N$ as the intrinsic carrier concentration *via* the standard relation [39]

$$N(T) = 2\left(\frac{2\pi kT}{h^2}\right)^{3/2} (m_n^* m_p^*)^{3/4} e^{-E_g/(2kT)} \quad (7)$$

where $k$ is Boltzmann's constant, $T$ is temperature, $h$ is the Planck constant, and $m_n^*$ and $m_p^*$ are the electron and hole effective masses, respectively. The relevant material properties for $Ga_{0.46}In_{0.54}As$ are taken from reference [40], and the temperature-dependent mobility is calculated according to reference [41]. With these equations, the complete temperature-dependent optical properties of $Ga_xIn_{1-x}As$ can be expressed as

$$\varepsilon(E) = \varepsilon'_{Adachi}(E) + i\left[\begin{array}{l}\varepsilon''_{Adachi}(E > E_g + 5\text{ meV}) + \\ \varepsilon''_{Guerra}(E \leq E_g + 5\text{ meV}) + \varepsilon''_{Drude}(E \leq E_g + 5\text{ meV})\end{array}\right] \quad (8)$$

The relative permittivity is then related to the refractive index and extinction coefficient by $\sqrt{\varepsilon} = n + ik$. The room temperature optical properties are shown in Fig. 3 for all layers in the absorber.

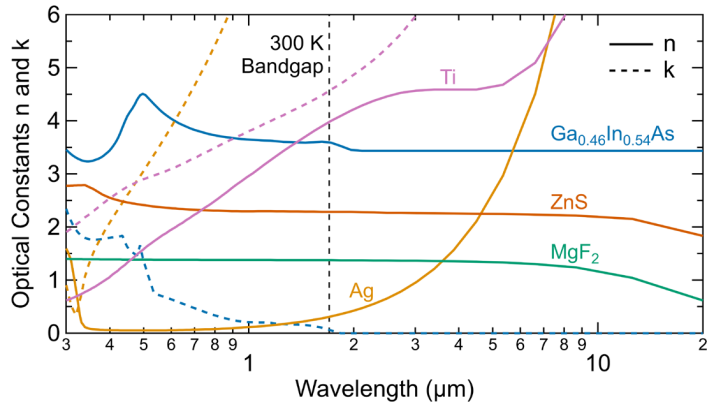

**Figure 3**. Optical properties used in the model for all layers in the selective solar absorber structure. The extinction coefficients for $MgF_2$ and ZnS are assumed to be zero in this range in accordance with the tabulated data.

**Results and Discussion**

The design of our selective solar absorber is influenced by our past work on high-quality $Ga_{0.46}In_{0.54}As$ thermophotovoltaic cells with rear reflectors that recycle sub-bandgap photons to the thermal emitter [42-44]. Because these share desirable optical characteristics with selective absorbers, we use the same material and approach in our absorber structure. In many ways, though, a selective absorber is a simpler structure than a thermophotovoltaic cell: no junction, electrical contacts, doping, nor cladding layers are needed. It is not even clear that single-crystal semiconductor material is necessary, only that it have a well-defined bandgap. These loose constraints allow us to grow only the undoped, actively absorbing $Ga_{0.46}In_{0.54}As$ semiconductor by metalorganic vapor phase epitaxy and deposit the other layers by standard evaporation techniques (Methods). It is important to note that the epitaxial growth process used here is intended to produce very high-quality crystalline photovoltaic cells and would be cost-prohibitive in a manufacturing setting for this type of selective absorber. However, low-cost semiconductor deposition methods such as sputtering could be used to scale up manufacturing of $Ga_xIn_{1-x}As$ films at much lower cost [45].

III-V semiconductors have a relatively large index of refraction (see Fig. 3), which leads to an undesirable solar reflectance of about 30% from an air-semiconductor interface. To mitigate this, we use a four-layer $MgF_2$/ZnS/$MgF_2$/ZnS antireflective coating (Fig. 2) commonly used for broadband solar absorption with multijunction photovoltaic cells [46, 47]. The effect of the antireflective coating can be seen in Fig. 4, which shows calculations of the spectral, angular absorptance or emittance for a simplified absorber with (Fig. 4a) and without (Fig. 4b) the antireflective coating. For wavelengths longer than the bandgap, waveguide modes at a semiconductor-metal interface can lead to infrared emissive losses [24-27]. These are reduced by the insertion of a thin spacer layer of $MgF_2$, whose effect can be seen by comparing the long wavelength region of Fig. 4a (no dielectric spacer) to Fig. 4b (with dielectric spacer). This figure also demonstrates that the absorptance and emittance characteristics are expected to weakly depend on angle of incidence up to large angles of about 75°, especially for wavelengths longer than the bandgap.

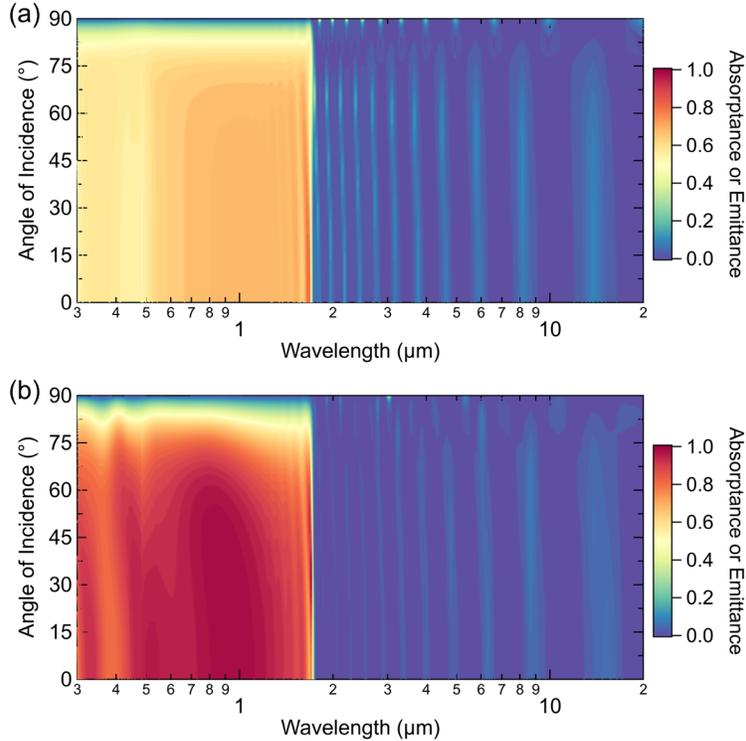

**Figure 4.** Calculated spectral, angular absorptance or emittance of (a) a simplified absorber structure with no antireflective coating and no dielectric spacer in between the semiconductor and rear reflector and (b) our selective absorber structure including those layers. The antireflective coating increases solar absorptance and the dielectric spacer reduces thermal emittance.

The optimum thickness of each layer in the absorber was determined by maximizing the solar-weighted absorptance while minimizing the blackbody-weighted infrared emittance predicted by our model with a Nelder-Mead simplex algorithm in MATLAB [48]. This yielded the layer thicknesses specified in Fig. 2a, and this structure was fabricated as described in the Methods section. The resulting absorber has a very uniform semiconductor layer, shown by cross-sectional scanning electron microscopy in Fig. 2b, which is expected from the metalorganic vapor phase epitaxy growth process. Some discontinuities in the antireflective coating can be seen in this image, but it is difficult to know if these result from nonideal evaporation of the dielectric layers or from damage during the focused ion beam cutting process used to expose the absorber cross-section. Nevertheless, the surface of our absorber is smooth and specular, as seen in the photograph of Fig. 2c.

The reflectance of the solar absorber was measured in the wavelength range of 300 nm – 20 µm while varying the temperature of the absorber from 30 °C to 100 °C (Methods), and the results (one minus reflectance) are shown in Fig. 5a along with the predictions of our model for each temperature, which overlap each other and appear as a single line away from the band edge. Our absorber shows high absorptance at solar wavelengths, low emittance at thermal wavelengths, and a remarkably sharp transition from high to low absorptance at about 1.75 µm, which is almost exactly the optimal transition wavelength shown in Fig. 1b. The results are insensitive to temperature in this range, with the exception of the expected bandgap shift to longer wavelengths as temperature is increased. Very good agreement with our model is observed across this spectral region, with the bandgap shift being captured particularly well by our calculations. Some deviations from the model are seen between 300 nm and 500 nm and for wavelengths greater than 10 μm, which likely results from as-deposited dielectric thicknesses differing from their nominal values. The small peaks near 3 μm are likely due to adsorbed water [49]. Typical interference fringes for a specular thin film are observed above and below the bandgap wavelength, although the magnitude of these fringes is small.

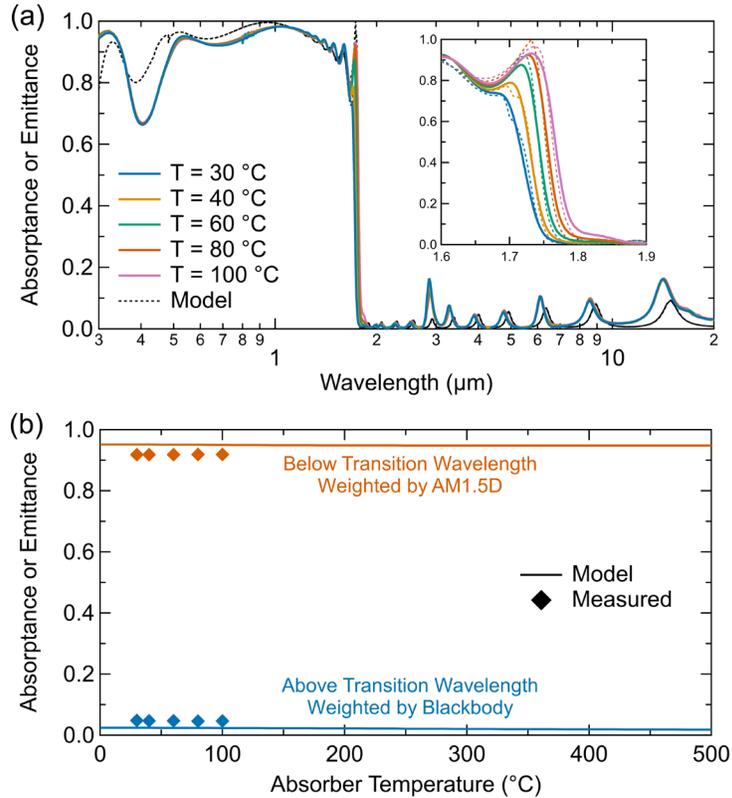

**Figure 5.** Measured and modeled temperature-dependent (a) near-normal absorptance and (b) below transition wavelength total solar absorptance weighted by the AM1.5D spectrum and above transition wavelength total hemispherical emittance calculated according to the absorber temperature. The absorber achieves high solar absorptance, low thermal emittance, and a sharp transition between these two regions.

Based on these measurements and our model, we calculate in Fig. 5b the weighted solar absorptance below the transition wavelength and the blackbody-weighted hemispherical thermal emittance above the transition wavelength. For the latter, we integrate over all angles of incidence with our model and assume the emittance is independent of angle for the experimental data, which is justified based on the very weak angular dependence predicted by our model and seen in Fig. 4. The measured solar absorptance is between 91.8% and 91.9% and the measured thermal emittance is between 4.6% and 4.8% below and above the transition wavelength, respectively, demonstrating that our structure is very effective both at absorbing solar energy and suppressing thermal emission. The model predicts that these numbers could be further improved to 95.0% absorptance and 2.4% emittance below and above the transition with improvement of layer quality and actual layer thickness. Importantly, the experimental data are stable with temperature over the tested range and the model indicates that these optical characteristics should hold at higher temperatures. The semiconductor's stability with temperature results from the fact that it is undoped and the intrinsic carrier concentration ($7.4 \times 10^{16}$ cm$^{-3}$ at 500 °C) does not rise enough to make free carrier emission problematic.

While these results are very promising, the true potential of our structure lies in the ability to tune the semiconductor composition for a particular transition wavelength that is specifically suited to an application's solar concentration and operating temperature. Our absorbers are epitaxially grown lattice-matched to an InP substrate (Methods), which resulted in a semiconductor composition of Ga$_{0.46}$In$_{0.54}$As and a 300 K bandgap energy of 0.73 eV (1.70 µm). The composition could be changed with the same growth method through lattice-mismatched epitaxy *via* a graded buffer [50-52] or, for a lower cost method, with a physical vapor deposition process such as sputtering [45]. To illustrate this potential, we calculate in Fig. 6 the spectral absorptance and emittance of our same absorber structure except for a semiconductor composition optimized for a particular solar concentration and absorber temperature. Fig. 6a shows a solar concentration of one with an absorber temperature of 250 °C, Fig. 6b shows a solar concentration of 50 with an absorber temperature of 500 °C, and Fig. 6c shows a solar concentration of 100 with an absorber temperature of 750 °C. The corresponding semiconductor compositions are Ga$_{0.49}$In$_{0.51}$As, Ga$_{0.57}$In$_{0.43}$As, and Ga$_{0.83}$In$_{0.17}$As with 300 K bandgaps of 0.76 eV (1.63 µm), 0.85 eV (1.46 µm), and 1.17 eV (1.06 µm), respectively.

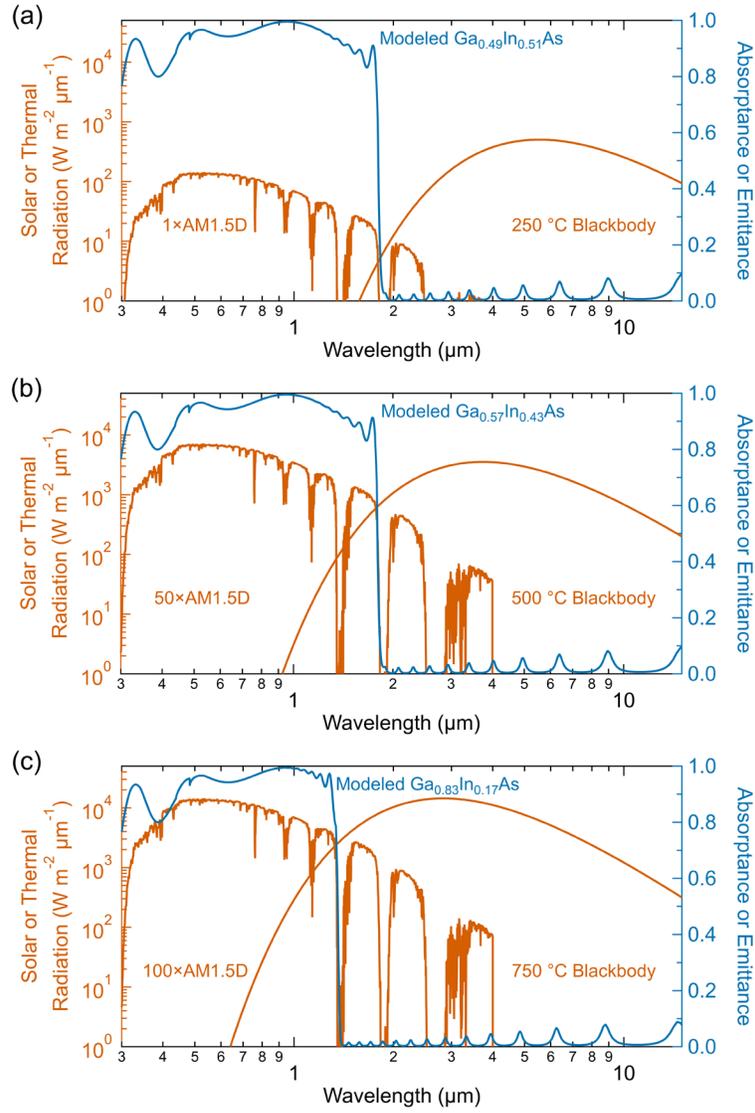

**Figure 6.** Modeled near-normal absorptance for solar absorbers with optimized semiconductor compositions for (a) a solar concentration of 1 and a temperature of 250 °C, (b) a solar concentration of 50 and a temperature of 500 °C, and (c) and solar concentration of 100 and a temperature of 750 °C.

## Conclusions

We have fabricated and characterized a semiconductor-based selective solar absorber that shows remarkable spectral selectivity at the bandgap wavelength of the semiconductor. The use of $Ga_{0.46}In_{0.54}As$ with an antireflective coating as the absorbing material provides excellent solar absorption for wavelengths shorter than the bandgap and very low absorption for longer wavelengths. Thermal emission is further suppressed with the use of a dielectric-metal rear reflector that improves upon the standard metal rear reflector in semiconductor-metal tandem selective absorbers. Importantly, the use of a ternary III-V semiconductor as the absorber allows the composition to be changed to tailor the absorber for a particular combination of solar concentration and operating temperature, which leads to high predicted solar-thermal efficiencies across a broad range of temperatures and solar concentrations.

Future work should focus on testing these absorbers at higher temperature to verify their stability under various atmospheric or vacuum conditions. Our absorbers included an epoxy layer that limited high temperature stability, but this could be replaced with a more robust interface such as a metal-metal cold weld bond. Another area for further investigation is alternate deposition methods for the semiconductor that could be much lower cost than the metalorganic vapor phase epitaxy used here. Because the electrical properties of the semiconductor are irrelevant for this application, a variety of other deposition techniques could be utilized that lead to polycrystalline material, as long as the semiconductor's optical properties are preserved.

The promising performance of our selective absorber structure strongly motivates further research in the development and understanding of semiconductor-dielectric-metal selective absorbers. With continued work on the manufacturing and optimization of this type of absorber, our structure could lead to a class of absorbers with near-perfect optical properties for any solar-thermal energy conversion process.

## Acknowledgements

This work was authored by the National Renewable Energy Laboratory (NREL), operated by Alliance for Sustainable Energy, LLC, for the U.S. Department of Energy (DOE) under Contract No. DE-AC36-08GO28308. This work was supported by the Laboratory Directed Research and Development (LDRD) Program at NREL. The views expressed in the article do not necessarily represent the views of the DOE or the U.S. Government. The U.S. Government retains and the publisher, by accepting the article for publication, acknowledges that the U.S. Government retains a nonexclusive, paid-up, irrevocable, worldwide license to publish or reproduce the published form of this work, or allow others to do so, for U.S. Government purposes.

## References

[1] R. Kempener, "Solar heat for industrial processes," IEA-ETSAP and IRENA Technology Brief E21, 2015.
[2] S. H. Farjana, N. Huda, M. A. P. Mahmud, and R. Saidur, "Solar process heat in industrial systems – A global review," *Renew. Sust. Energ. Rev.,* vol. 82, pp. 2270-2286, 2018, doi: 10.1016/j.rser.2017.08.065.

[3] L. A. Weinstein, J. Loomis, B. Bhatia, D. M. Bierman, E. N. Wang, and G. Chen, "Concentrating Solar Power," *Chem. Rev.,* vol. 115, no. 23, pp. 12797-12838, 2015, doi: 10.1021/acs.chemrev.5b00397.

[4] C. Augustine, C. Turchi, and M. Mehos, "The role of concentrating solar-thermal technologies in a decarbonized U.S. grid," National Renewable Energy Laboratory, Golden, CO, 2021. [Online]. Available: https://www.nrel.gov/docs/fy21osti/80574.pdf

[5] C. E. Kennedy, "Review of Mid- to High-Temperature Solar Selective Absorber Materials," National Renewable Energy Laboratory, United States, NREL/TP-520-31267, 2002. [Online]. Available: https://www.osti.gov/biblio/15000706

[6] P. Bermel, J. Lee, J. D. Joannopoulos, I. Celanovic, and M. Soljačić, "Selective solar absorbers," *Annu. Rev. Heat Transfer,* vol. 15, pp. 231-254, 2012, doi: 10.1615/AnnualRevHeatTransfer.2012004119.

[7] V. Rinnerbauer *et al.*, "Recent developments in high-temperature photonic crystals for energy conversion," *Energy Environ. Sci.,* 10.1039/C2EE22731B vol. 5, no. 10, pp. 8815-8823, 2012, doi: 10.1039/C2EE22731B.

[8] N. Selvakumar and H. C. Barshilia, "Review of physical vapor deposited (PVD) spectrally selective coatings for mid- and high-temperature solar thermal applications," *Sol. Energy Mater. Sol. Cells,* vol. 98, pp. 1-23, 2012, doi: 10.1016/j.solmat.2011.10.028.

[9] F. Cao, K. McEnaney, G. Chen, and Z. Ren, "A review of cermet-based spectrally selective solar absorbers," *Energy Environ. Sci.,* vol. 7, no. 5, pp. 1615-1627, 2014, doi: 10.1039/C3EE43825B.

[10] A. Amri, Z. T. Jiang, T. Pryor, C.-Y. Yin, and S. Djordjevic, "Developments in the synthesis of flat plate solar selective absorber materials via sol–gel methods: A review," *Renew. Sust. Energ. Rev.,* vol. 36, pp. 316-328, 2014, doi: 10.1016/j.rser.2014.04.062.

[11] M. Bello and S. Shanmugan, "Achievements in mid and high-temperature selective absorber coatings by physical vapor deposition (PVD) for solar thermal Application-A review," *Journal of Alloys and Compounds,* vol. 839, p. 155510, 2020, doi: 10.1016/j.jallcom.2020.155510.

[12] K. Xu, M. Du, L. Hao, J. Mi, Q. Yu, and S. Li, "A review of high-temperature selective absorbing coatings for solar thermal applications," *Journal of Materiomics,* vol. 6, no. 1, pp. 167-182, 2020, doi: 10.1016/j.jmat.2019.12.012.

[13] D. K. Edwards, J. T. Gier, K. E. Nelson, and R. D. Roddick, "Spectral and directional thermal radiation characteristics of selective sufaces for solar collectors," *Solar Energy,* vol. 6, no. 1, pp. 1-8, 1962, doi: 10.1016/0038-092X(62)90092-0.

[14] D. A. Williams, T. A. Lappin, and J. A. Duffie, "Selective radiation properties of particulate coatings," *Journal of Engineering for Power,* vol. 85, no. 3, pp. 213-220, 1963, doi: 10.1115/1.3675262.

[15] T. J. McMahon and D. L. Stierwalt, "Cost-effective PbS-Al selective solar-absorbing panel," in *SPIE 0068, Optics in Solar Energy Utilization*, 1976, doi: 10.1117/12.978116.

[16] S. Chatterjee and U. Pal, "Low-cost solar selective absorbers from Indian galena," *Optical Engineering,* vol. 32, no. 11, pp. 2923-2929, 1993, doi: 10.1117/12.148123.

[17] B. O. Seraphin, "Chemical vapor deposition of thin semiconductor films for solar energy conversion," *Thin Solid Films,* vol. 39, pp. 87-94, 1976, doi: 10.1016/0040-6090(76)90626-X.

[18] A. Donnadieu and B. O. Seraphin, "Optical performance of absorber-reflector combinations for photothermal solar energy conversion," *J. Opt. Soc. Am.,* vol. 68, no. 3, pp. 292-297, 1978, doi: 10.1364/JOSA.68.000292.

[19] D. C. Booth, D. D. Allred, and B. O. Seraphin, "Stabilized CVD amorphous silicon for high temperature photothermal solar energy conversion," *Solar Energy Materials,* vol. 2, no. 1, pp. 107-124, 1979, doi: 10.1016/0165-1633(79)90034-0.

[20] M. Okuyama, K. Saji, T. Adachi, H. Okamoto, and Y. Hamakawa, "Selective absorber using glow-discharge amorphous silicon for solar photothermal conversion," *Solar Energy Materials,* vol. 3, no. 3, pp. 405-413, 1980, doi: 10.1016/0165-1633(80)90029-5.

[21] P. Bermel *et al.*, "Design and global optimization of high-efficiency thermophotovoltaic systems," *Opt. Express,* vol. 18, no. S3, pp. A314-A334, 2010, doi: 10.1364/OE.18.00A314.

[22] N. H. Thomas, Z. Chen, S. Fan, and A. J. Minnich, "Semiconductor-based Multilayer Selective Solar Absorber for Unconcentrated Solar Thermal Energy Conversion," *Sci. Rep.,* vol. 7, no. 1, p. 5362, 2017, doi: 10.1038/s41598-017-05235-x.

[23] J. Moon *et al.*, "High performance multi-scaled nanostructured spectrally selective coating for concentrating solar power," *Nano Energy,* vol. 8, pp. 238-246, 2014, doi: 10.1016/j.nanoen.2014.06.016.

[24] R. M. Swanson, "Recent developments in thermophotovoltaic conversion," in *1980 International Electron Devices Meeting*, 1980, pp. 186-189, doi: 10.1109/IEDM.1980.189789.

[25] T. Burger, D. Fan, K. Lee, S. R. Forrest, and A. Lenert, "Thin-Film Architectures with High Spectral Selectivity for Thermophotovoltaic Cells," *ACS Photonics,* vol. 5, no. 7, pp. 2748-2754, 2018, doi: 10.1021/acsphotonics.8b00508.

[26] D. Fan, T. Burger, S. McSherry, B. Lee, A. Lenert, and S. R. Forrest, "Near-perfect photon utilization in an air-bridge thermophotovoltaic cell," *Nature,* vol. 586, no. 7828, pp. 237-241, 2020, doi: 10.1038/s41586-020-2717-7.

[27] M. K. Arulanandam *et al.*, "Rigorous Coupled Wave Analysis of GaAs Thermophotovoltaic Devices with a Patterned Dielectric Back Contact," in *2021 IEEE 48th Photovoltaic Specialists Conference (PVSC)*, 20-25 June 2021 2021, pp. 2580-2583, doi: 10.1109/PVSC43889.2021.9518396.

[28] P. Yeh, *Optical Waves in Layered Media*. Hoboken, New Jersey: John Wiley & Sons, Inc., 2005.

[29] H. U. Yang, J. D'Archangel, M. L. Sundheimer, E. Tucker, G. D. Boreman, and M. B. Raschke, "Optical dielectric function of silver," *Phys. Rev. B,* vol. 91, no. 23, p. 235137, 2015, doi: 10.1103/PhysRevB.91.235137.

[30] A. D. Rakić, A. B. Djurišić, J. M. Elazar, and M. L. Majewski, "Optical properties of metallic films for vertical-cavity optoelectronic devices," *Appl. Opt.,* vol. 37, no. 22, pp. 5271-5283, 1998, doi: 10.1364/AO.37.005271.

[31] M. J. Dodge, "Refractive properties of magnesium fluoride," *Appl. Opt.,* vol. 23, no. 12, pp. 1980-1985, 1984, doi: 10.1364/AO.23.001980.

[32] M. R. Querry, "Optical Constants of Minerals and Other Materials from the Millimeter to the Ultraviolet," U.S. Army Chemical Research Development and Engineering Center, Aberdeen Proving Ground, Maryland, ADA192210, 1987.

[33] S. Adachi, "Optical dispersion relations for GaP, GaAs, GaSb, InP, InAs, InSb, $Al_xGa_{1-x}As$, and $In_{1-x}Ga_xAs_yP_{1-y}$," *J. Appl. Phys.,* vol. 66, no. 12, pp. 6030-6040, 1989, doi: 10.1063/1.343580.

[34] E. Zielinski, H. Schweizer, K. Streubel, H. Eisele, and G. Weimann, "Excitonic transitions and exciton damping processes in InGaAs/InP," *J. Appl. Phys.,* vol. 59, no. 6, pp. 2196-2204, 1986, doi: 10.1063/1.336358.

[35] I. Vurgaftman, J. R. Meyer, and L. R. Ram-Mohan, "Band parameters for III–V compound semiconductors and their alloys," *J. Appl. Phys.,* vol. 89, no. 11, pp. 5815-5875, 2001, doi: 10.1063/1.1368156.

[36] L. Vegard, "Die Konstitution der Mischkristalle und die Raumfüllung der Atome," *Zeitschrift für Physik,* vol. 5, no. 1, pp. 17-26, 1921, doi: 10.1007/BF01349680.

[37] J. A. Guerra, A. Tejada, J. A. Töfflinger, R. Grieseler, and L. Korte, "Band-fluctuations model for the fundamental absorption of crystalline and amorphous semiconductors: a dimensionless joint density of states analysis," *Journal of Physics D: Applied Physics,* vol. 52, no. 10, p. 105303, 2019, doi: 10.1088/1361-6463/aaf963.

[38] S. Basu, B. J. Lee, and Z. M. Zhang, "Infrared Radiative Properties of Heavily Doped Silicon at Room Temperature," *Journal of Heat Transfer,* vol. 132, no. 2, 2009, doi: 10.1115/1.4000171.

[39] D. A. Neamen, *Semiconductor Physics and Devices: Basic Principles*, 4 ed. New York: McGraw-Hill, 2012, p. 784.

[40] M. Levinshtein, S. Rumyantsev, and M. Shur, *Handbook Series on Semiconductor Parameters*. WORLD SCIENTIFIC, 1996, p. 452.

[41] M. Sotoodeh, A. H. Khalid, and A. A. Rezazadeh, "Empirical low-field mobility model for III–V compounds applicable in device simulation codes," *J. Appl. Phys.,* vol. 87, no. 6, pp. 2890-2900, 2000, doi: 10.1063/1.372274.

[42] Z. Omair *et al.*, "Ultraefficient thermophotovoltaic power conversion by band-edge spectral filtering," *Proc. Natl. Acad. Sci. U. S. A.,* vol. 116, no. 31, p. 15356, 2019, doi: 10.1073/pnas.1903001116.

[43] T. C. Narayan *et al.*, "World record demonstration of > 30% thermophotovoltaic conversion efficiency," in *2020 47th IEEE Photovoltaic Specialists Conference (PVSC)*, 15 June-21 Aug. 2020 2020, pp. 1792-1795, doi: 10.1109/PVSC45281.2020.9300768.

[44] T. C. Narayan *et al.*, "Platform for Accurate Efficiency Quantification of > 35% Efficient Thermophotovoltaic Cells," in *2021 IEEE 48th Photovoltaic Specialists Conference (PVSC)*, 20-25 June 2021 2021, pp. 1352-1354, doi: 10.1109/PVSC43889.2021.9518588.

[45] R. Bernal-Correa, S. Gallardo-Hernández, J. Cardona-Bedoya, and A. Pulzara-Mora, "Structural and optical characterization of GaAs and InGaAs thin films deposited by RF magnetron sputtering," *Optik,* vol. 145, pp. 608-616, 2017, doi: 10.1016/j.ijleo.2017.08.042.

[46] D. J. Aiken, "High performance anti-reflection coatings for broadband multi-junction solar cells," *Sol. Energy Mater. Sol. Cells,* vol. 64, no. 4, pp. 393-404, 2000, doi: 10.1016/S0927-0248(00)00253-1.

[47] J. F. Geisz *et al.*, "Six-junction III–V solar cells with 47.1% conversion efficiency under 143 Suns concentration," *Nature Energy,* vol. 5, no. 4, pp. 326-335, 2020, doi: 10.1038/s41560-020-0598-5.

[48] J. C. Lagarias, J. A. Reeds, M. H. Wright, and P. E. Wright, "Convergence Properties of the Nelder--Mead Simplex Method in Low Dimensions," *SIAM J. Optimization,* vol. 9, no. 1, pp. 112-147, 1998, doi: 10.1137/S1052623496303470.

[49] NIST Mass Spectrometry Data Center, "Infrared Spectra," in *NIST Chemistry WebBook, NIST Standard Reference Database Number 69*, P. J. Linstrom and W. G. Mallard Eds. Gaithersburg MD: National Institute of Standards and Technology, 2021.

[50] J. F. Geisz *et al.*, "High-efficiency GaInP∕GaAs∕InGaAs triple-junction solar cells grown inverted with a metamorphic bottom junction," *Appl. Phys. Lett.,* vol. 91, no. 2, p. 023502, 2007, doi: 10.1063/1.2753729.

[51] J. F. Geisz *et al.*, "40.8% efficient inverted triple-junction solar cell with two independently metamorphic junctions," *Appl. Phys. Lett.,* vol. 93, no. 12, p. 123505, 2008, doi: 10.1063/1.2988497.

[52] R. M. France, F. Dimroth, T. J. Grassman, and R. R. King, "Metamorphic epitaxy for multijunction solar cells," *MRS Bulletin,* vol. 41, no. 3, pp. 202-209, 2016, doi: 10.1557/mrs.2016.25.

[53] J. A. Duffie and W. A. Beckman, *Solar Engineering of Thermal Processes*. Hoboken, New Jersey: John Wiley & Sons, Inc., 2013.